\documentclass[pdflatex,sn-mathphys-num]{sn-jnl}

\usepackage{graphicx}%
\usepackage{multirow}%
\usepackage{amsmath,amssymb,amsfonts}%
\usepackage{amsthm}%
\usepackage{mathrsfs}%
\usepackage[title]{appendix}%

\usepackage{textcomp}%
\usepackage{manyfoot}%
\usepackage{booktabs}%
\usepackage{algorithm}%
\usepackage{algorithmicx}%
\usepackage{algpseudocode}%
\usepackage{listings}%
\usepackage{caption}

\usepackage[table]{xcolor}
\usepackage{float}

\usepackage{longtable} 
\usepackage{enumitem}
\usepackage{epstopdf}      
\usepackage{subcaption} 
\usepackage{siunitx}
\theoremstyle{thmstyleone}%
\theoremstyle{thmstyletwo}%

\theoremstyle{thmstylethree}%

\begin{document}

\title[Article Title]{When Normality Tests Detect Equilibrium Distributions of Finite N-Body Systems}

\author[1,2,3]{\fnm{Jae Wan} \sur{Shim}}\email{jae-wan.shim@kist.re.kr}

\affil[1]{\orgdiv{Extreme Materials Research Center}, \orgname{Korea Institute of Science and Technology}, \orgaddress{\street{5 Hwarang-ro 14-gil, Seongbuk}, \city{Seoul}, \postcode{02792},  \country{Republic of Korea}}}

\affil[2]{\orgdiv{Climate and Environmental Research Institute}, \orgname{Korea Institute of Science and Technology}, \orgaddress{\street{5 Hwarang-ro 14-gil, Seongbuk}, \city{Seoul}, \postcode{02792},  \country{Republic of Korea}}}

\affil[3]{\orgdiv{Division of AI-Robotics}, \orgname{KIST Campus, University of Science and Technology}, \orgaddress{\street{5 Hwarang-ro 14-gil, Seongbuk}, \city{Seoul}, \postcode{02792},  \country{Republic of Korea}}}

\abstract{The particle number $N$ can be used as a quantitative gauge of non-Gaussianity. This idea extends to systems that are not literally finite by assigning them a notional $N$ that captures the same deviation. For an ideal gas with $N$ insufficiently large for the thermodynamic limit, the velocity distribution that maximises Havrda--Charv\'at entropy departs markedly from the Maxwell--Boltzmann (Gaussian) form obtained in that limit. We explore how five standard normality tests---Kolmogorov--Smirnov, Anderson--Darling, Cram\'er--von Mises, Jarque--Bera and Shapiro--Wilk---respond to samples drawn from this finite-$N$ equilibrium distribution. A large-scale Monte Carlo study maps the tests' statistical power across system size $N$ and sample size $n$, providing practical reference tables and a heuristic scaling law, visualised as a contour plot, that together indicate when finite-size effects remain detectable.}


\maketitle

\section{Introduction}

The Gaussian (Maxwell--Boltzmann) distribution is often invoked as a universal equilibrium state, its dominance justified by the central limit theorem and by kinetic arguments.  
When the number of degrees of freedom is \emph{finite}, however, the assumptions behind these asymptotic results break down: the exact most-probable distributions become markedly non-Gaussian, although they still converge smoothly to the Gaussian form as $N\!\rightarrow\!\infty$ \cite{shim2020entropy, shim2021minimal}.  Such finite-$N$ effects are not an academic curiosity; they appear in diverse contexts ranging from rarefied-gas flows to socioeconomic models \cite{Tsallis1988,Tsallis2009,Cho2002,Plastino1993,Plastino1994,TsallisStariolo1996,TsallisBemskiMendes1999,WeinsteinLloydTsallis2002,LiuGoree2008,WaltonRafelski2000,IonIon2001,Upadhyaya2001,Lutz2003,ReynoldsVeneziani2004,Reis2006,Beck2000,Borland2002,AusloosPetroni2007}.

In this work we ask when finite‐size deviations become statistically
indistinguishable from a Gaussian as \(N\) grows, using standard
normality tests as the diagnostic tool.  
By charting the statistical power of these tests on a grid of particle
numbers \(N\) and sample sizes \(n\), we provide an \emph{experimental
reference map} that helps researchers choose an adequate sample size and interpret the resulting $p$-values.  
Because normality tests compare the data directly with a fixed Gaussian
benchmark, this approach sidesteps the density-estimation step required
by information-theoretic measures.

The Havrda--Charv\'at entropy \cite{Havrda1967} provides a link between the particle number \(N\) of a system and the equilibrium state's departure from Gaussianity.  
For an ideal-gas ensemble of \(N\) identical particles in \(D\) spatial
dimensions, maximising this entropy subject to a fixed total kinetic
energy yields the compact-support equilibrium density \cite{shim2020entropy}
\begin{equation}
f(v)=\frac{\Gamma\bigl(\frac{DN}{2}\bigr)}
          {\Gamma\bigl(\frac{D(N-1)}{2}\bigr)\,(\pi U)^{D/2}}
          \Bigl(1-\frac{v^{2}}{U}\Bigr)^{\frac{D(N-1)-2}{2}}_{+},
\label{eq:finiteNpdf}
\end{equation}
valid for $N>1+2/D$. Here, $\Gamma(\,\cdot\,)$ denotes Euler's gamma function, and $(x)_{+}\!=\!\max(0,x)$, which restricts the domain to the compact support $\|v\| \le \sqrt{U}$. The term $mU/2$ denotes the total kinetic energy of the $N$-particle system where $m$ is the mass of a particle and $v$ is a $D$-dimensional velocity vector. The relation
\[
q=\frac{D(N-1)-4}{D(N-1)-2},
\]
first derived in Ref.~\cite{shim2020entropy}, endows the entropic index
\(q\) with a concrete physical meaning.  
Substituting this expression into Eq.~\eqref{eq:finiteNpdf} recasts the
density as a $q$-Gaussian.

The choice of the Havrda--Charv\'at entropy,
\[
  S_\alpha=\frac{1}{2^{1-\alpha}-1}\!\bigl(\sum_i p_i^{\alpha}-1\bigr),
\]
which is functionally identical to the Tsallis \(q\)-entropy \cite{Havrda1967,Tsallis1988}, is not arbitrary. 
In his derivation of the entropy formula, Boltzmann indispensably assumed an infinite number of particles to use Stirling's approximation, which leads to the Gaussian distribution. Maximising the Havrda--Charv\'at entropy under a fixed‐energy constraint naturally yields the compact‐support density in Eq.~\eqref{eq:finiteNpdf}, as rigorously proven in Ref.~\cite{shim2020entropy}.

The above derivation is further reinforced from a microcanonical perspective. 
The system state is uniformly distributed on the constant‐energy hypersphere according to the fundamental microcanonical assumption. By integrating the coordinates of \(N\!-\!1\) particles, we obtain the single-particle distribution which reproduces exactly the polynomial form of Eq.~\eqref{eq:finiteNpdf} \cite{shim2020entropy}. 

This paper studies how standard goodness-of-fit and normality tests \cite{d1986goodness} perform when data are drawn from the one-dimensional compact-support distribution of Eq.~\eqref{eq:finiteNpdf}. For each $N$ we generate independent samples and
\begin{enumerate}
\item assess type-I error control of the Kolmogorov--Smirnov and Anderson--Darling tests under the correctly specified (custom) null, and
\item evaluate the power of Kolmogorov--Smirnov, Anderson--Darling, Cram\'er--von Mises, Jarque--Bera, and Shapiro--Wilk tests to reject the null hypothesis of normality \cite{shapiro1965, jarque1987}.
\end{enumerate}

Because the generating law approaches Gaussian as $N$ increases, the experiment traces a controlled path from clearly non-Gaussian to nearly Gaussian regimes.  Well-calibrated tests should maintain nominal size against the true distribution for all $N$, while their power against normality should decay as the underlying law becomes more Gaussian. This is a classic trade-off in hypothesis testing between the probability of a type-I error (size) and that of a type-II error (related to power) \cite{lehmann2005}.

In the following sections, we quantify, in a principled finite-$N$ setting, how nominal size and power of widely used tests evolve with $N$. We identify which tests are most responsive to the short-tailed deviations characteristic of compact-support $q$-Gaussians and indicate how quickly normality assumptions become adequate as $N$ grows. Such power comparisons are crucial for providing practical guidance to researchers \cite{razali2011}.

\section{Simulation Procedure} \label{sec:simulation}

\subsection{Target distribution}

For a one-dimensional system ($D=1$) with a finite particle count $N$, the code defines the entropic index
\[
q(D,N)=\frac{D(N-1)-4}{D(N-1)-2},\qquad\text{so here } q_N\equiv q(1,N)=\frac{N-5}{N-3}.
\]
Using this value, a compact-support \emph{finite-$N$ distribution} is defined by the density
\[
p_{N,U}(x)=
\frac{\Gamma\bigl(\frac{2-q_N}{1-q_N}+\frac{1}{2}\bigr)}
     {\Gamma\bigl(\frac{2-q_N}{1-q_N}\bigr)\,(\pi U)^{1/2}}
     \Bigl(1-\frac{x^{2}}{U}\Bigr)_{+}^{\frac{1}{1-q_N}},
\]
where we set \(U=N\) for every experiment.  
This choice reflects the physical extensivity of the total kinetic
energy---energy grows linearly with the particle number---and at the same
time fixes the second moment of \(p_{N,U}\) to unity for all
\(N>3\).  
With the variance normalised in this way, the finite-\(N\) density can
be compared to the standard Gaussian \(\mathcal N(0,1)\) without any
additional rescaling.  
The operator \((\cdot)_{+}\) denotes the positive part, so the density
vanishes outside the compact support
\(x\in[-\sqrt{U},\sqrt{U}]\).

\subsection{Loop structure}

\begin{description}[leftmargin=2.5em, labelsep=0.5em]

\item[Outer loop (system size).]  
$N$ takes integer values from $5$ to $20$ inclusive:
\[
N \in \{5,6,\dots,20\}.
\]

\item[Sample-size loop.]  
Define the baseline grid of sample sizes:
\[
\begin{split}
  \mathcal{S} = \{ & 10, 20, 30, 40, 50, 60, 70, 80, 90, 100, 500, 1\,000, \\
                   & 2\,000, 3\,000, 4\,000, 5\,000, 10\,000, 20\,000, 50\,000 \}.
\end{split}
\]

\begin{itemize}
  \item For $N = 5,\dots,19$, we use the set \(n \in \mathcal{S}\).
  \item For $N = 20$, the set is extended to include \(n = 100\,000\).
\end{itemize}

\item[Monte Carlo loop.]  
For every $(N,n)$ pair, the code performs $100$ independent replicates. In each iteration, a fresh set of $n$ samples is drawn using the built-in random number generator.
\begin{enumerate}[label=\alph*)]
   \item Draw $n$ i.i.d.\ variates from the custom $q$-Gaussian. Given the compact support and the bounded nature of the density, rejection sampling is employed to ensure exact generation.
       \item Apply the Kolmogorov--Smirnov (KS) and Anderson--Darling (AD) tests \emph{under the correct null} (the custom density) and count a ``success'' if the null is \emph{not} rejected at $\alpha=0.05$ (type-I control).
    \item Apply KS, AD, and Cram\'er--von Mises (CvM) tests \emph{against the null hypothesis of normality}; apply Jarque--Bera (JB) and Shapiro--Wilk (SW) classical normality tests.  
          A ``success'' is recorded when the test \emph{does} reject normality at the same $\alpha=0.05$ (empirical power).
\end{enumerate}
Seven counters are recorded for every \((N,n)\):
\[
\bigl(\text{KS}_{\text{cust}},\,
      \text{AD}_{\text{cust}},\,
      \text{KS},\,
      \text{AD},\,
      \text{CvM},\,
      \text{JB},\,
      \text{SW}\bigr),
\]
where
\begin{itemize}
    \item \(\text{KS}_{\text{cust}}\) and \(\text{AD}_{\text{cust}}\) tally \emph{Type-I control}: the count increases when the KS or AD test \emph{fails to reject} the \emph{true} finite-\(N\) distribution at \(\alpha=0.05\).
    \item The remaining five counters (\(\text{KS},\ \text{AD},\ \text{CvM},\ \text{JB},\ \text{SW}\)) measure \emph{power against normality}: each is incremented when the corresponding test \emph{does} reject the null hypothesis of Gaussianity at the same significance level.
\end{itemize}

For the assessment of Type-I error control, the null hypothesis was fully specified ($N$ and $U=N$ fixed). Consequently, the critical values were obtained via numerical inversion of the distribution function.
\end{description}

\section{Simulation results}\label{sec:results}

We summarise the empirical findings in
Table~\ref{tab:summary_power} (selected $N,n$ grid) and
Table~\ref{tab:sample_size_scan} ($N=20$ size-scan). The complete dataset covering the full $(N,n)$ grid is provided in Table~\ref{tab:merged_power_map} in the Appendix.
Three questions guide the discussion:
\textit{(i)} Do goodness-of-fit tests respect the nominal
$\alpha=0.05$ size?
\textit{(ii)} How does power vary with the finite-$N$ parameter?
\textit{(iii)} How much data are required once the distribution is
nearly Gaussian?

\begin{table}[t!]
\centering
\caption{Summary of statistical power for selected sample sizes. For each system size $N$, we report the power at small sample sizes ($n=10, 50$), the transition point where tests sensitive to moment deviations reach ${\sim}50\%$ power, and the saturation point where they reach ${\sim}100\%$ power. (Full results are provided in Appendix Table~\ref{tab:merged_power_map}.)}
\label{tab:summary_power}
\small
\begin{tabular}{cc ccccc}
\toprule
\textbf{Param.} & \textbf{Sample} & \multicolumn{5}{c}{\textbf{Statistical Power (\%)}} \\
\cmidrule(lr){3-7}
$\boldsymbol{N}$ & \textbf{Size} $\boldsymbol{n}$ & \textbf{KS} & \textbf{AD} & \textbf{CvM} & \textbf{JB} & \textbf{SW} \\
\midrule

\multirow{4}{*}{5} 
 & 10   & 5 & 7 & 6 & 1 & 4 \\
 & 50   & 8 & 7 & 7 & 0 & 16 \\
 & 100  & 10 & 7 & 6 & 2 & 44 \\
 & 500  & 40 & 49 & 30 & 100 & 100 \\
\midrule

\multirow{4}{*}{8} 
 & 10   & 8 & 9 & 8 & 6 & 5 \\
 & 50   & 8 & 11 & 9 & 0 & 1 \\
 & 500  & 16 & 16 & 16 & 73 & 84 \\
 & \num{1000} & 30 & 29 & 19 & 100 & 100 \\
\midrule

\multirow{4}{*}{11} 
 & 10   & 5 & 3 & 3 & 3 & 5 \\
 & 50   & 2 & 3 & 3 & 0 & 4 \\
 & 500  & 9 & 8 & 7 & 42 & 48 \\
 & \num{2000} & 31 & 28 & 17 & 100 & 100 \\
\midrule

\multirow{4}{*}{14} 
 & 10   & 9 & 10 & 10 & 4 & 2 \\
 & 50   & 8 & 8 & 10 & 0 & 4 \\
 & \num{1000} & 10 & 11 & 9 & 59 & 61 \\
 & \num{3000} & 19 & 21 & 14 & 98 & 97 \\
\midrule

\multirow{4}{*}{17} 
 & 10   & 5 & 5 & 4 & 3 & 3 \\
 & 50   & 4 & 6 & 5 & 2 & 7 \\
 & \num{2000} & 16 & 13 & 13 & 74 & 72 \\
 & \num{5000} & 23 & 28 & 20 & 100 & 100 \\
\midrule

\multirow{4}{*}{19} 
 & 10   & 1 & 4 & 3 & 2 & 3 \\
 & 50   & 6 & 5 & 6 & 1 & 1 \\
 & \num{2000} & 9 & 9 & 8 & 65 & 59 \\
 & \num{5000} & 22 & 24 & 20 & 100 & 99 \\

\bottomrule
\end{tabular}
\end{table}

\begin{table}
\centering
\caption{Extended analysis of statistical power as a function of sample size for the custom distribution with parameter $N=20$. The validation columns have been omitted to focus on the power comparison.}
\label{tab:sample_size_scan}
\begin{tabular}{cccccc}
\toprule
\textbf{Sample Size} & \multicolumn{5}{c}{\textbf{Statistical Power (\%)}} \\
\cmidrule(lr){2-6}
$\boldsymbol{n}$ & \textbf{KS} & \textbf{AD} & \textbf{C-vM} & \textbf{JB} & \textbf{SW} \\
\midrule
10      & 9  & 11 & 9  & 3   & 7   \\
20      & 9  & 7  & 7  & 3   & 8   \\
30      & 5  & 7  & 7  & 4   & 6   \\
40      & 5  & 5  & 5  & 2   & 0   \\
50      & 7  & 9  & 8  & 2   & 1   \\
60      & 4  & 4  & 4  & 1   & 5   \\
70      & 1  & 5  & 4  & 0   & 5   \\
80      & 6  & 8  & 7  & 0   & 4   \\
90      & 2  & 3  & 5  & 2   & 6   \\
100     & 2  & 3  & 4  & 0   & 7   \\
500     & 3  & 3  & 4  & 7   & 14  \\
\num{1000}   & 11 & 12 & 9  & 19  & 21  \\
\num{2000}   & 13 & 10 & 9  & 51  & 50  \\
\num{3000}   & 14 & 13 & 6  & 84  & 80  \\
\num{4000}   & 20 & 24 & 16 & 97  & 86  \\
\num{5000}   & 23 & 27 & 21 & 98  & 96  \\
\num{10000}  & 31 & 48 & 29 & 100 & 100 \\
\num{20000}  & 74 & 91 & 73 & 100 & 100 \\
\num{50000}  & 99 & 100& 100& 100 & 100 \\
\num{100000} & 100& 100& 100& 100 & 100 \\
\bottomrule
\end{tabular}
\end{table}

\subsection{Size control under the custom null}

For every $(N,n)$ pair the Kolmogorov--Smirnov
(KS\textsubscript{cust}) and Anderson--Darling
(AD\textsubscript{cust}) columns remain close to the
$95\%$ non-rejection rate, confirming reliable type-I control.

\subsection{Power as a function of $N$}

Reading Table~\ref{tab:merged_power_map} downwards shows a monotone
decline in power as $N$ rises:

\begin{itemize}
\item For strongly non-Gaussian cases ($5\le N\le 9$)  
      Jarque--Bera (JB) and Shapiro--Wilk (SW) reject normality in
      essentially every trial even at $n=500$,
      while empirical cumulative distribution function (ECDF)-based tests (KS, AD, Cram\'er--von Mises or CvM) reach
      at most ${\sim}50\%$.
\item Once $N\ge 14$ the ECDF tests rarely exceed $20\%$ power
      unless $n$ is in the thousands, whereas JB and SW still exceed
      $60\%$ for the same sample size.
\end{itemize}

Hence, excess kurtosis provides a more persistent signal of non-normality than the global CDF shape. This observation aligns with the theoretical framework of the D'Agostino--Pearson $K^2$ test \cite{d1973tests}.
\subsection{Sample-size effects at $N=20$}

Table~\ref{tab:sample_size_scan} isolates the near-Gaussian case
$N=20$.

\begin{enumerate}[label=\textbf{\arabic*.}, leftmargin=*]

\item \textbf{Shape-based tests need more data.}
      KS, AD and CvM start below $30\%$ at $n=\num{5000}$ and only reach
      ${\sim}100\%$ at $n\ge \num{50000}$.
\item \textbf{Tests sensitive to moment deviations are efficient.}
      JB and SW already exceed $95\%$ at $n=5\,000$ and hit
      $100\%$ by $n=10\,000$.
\end{enumerate}

\subsection{Summary}

\begin{enumerate}[label=\roman*), leftmargin=*]
\item \textbf{Type-I control} is satisfactory for KS and AD under the
      correctly specified null.
\item \textbf{Finite-$N$ convergence.}
      All tests lose power as $N\to\infty$, mirroring the theoretical
      approach to Gaussian form.
\item \textbf{Test hierarchy.}
      Tests sensitive to moment deviations (JB, SW) dominate ECDF-based ones
      (KS, AD, CvM) across most of the $(N,n)$ plane.
\item \textbf{Practical guidance.}
      When the distribution may differ from normality mainly
      through higher-order moments, JB or SW should be preferred;
      ECDF tests demand much larger samples to reach comparable
      power.
\end{enumerate}

\subsection{Theoretical Interpretation and Scaling Heuristic}
\label{sec:scaling}
The superior performance of tests sensitive to moment deviations (JB, SW) over ECDF-based tests (KS, AD, CvM) can be understood through the moment structure of the finite-$N$ distribution. 
Since the distribution is symmetric, its skewness is identically zero. The deviation from Gaussianity is thus concentrated in the \emph{kurtosis}. 
Explicit calculation of the moments for this microcanonical ensemble yields the exact excess kurtosis $\kappa_{\text{ex}}$:\footnote{For the density $f(v) \propto (1-v^2/N)^{(N-3)/2}$, the moments are evaluated using standard Beta function integrals. With the normalization $\langle v^2 \rangle = 1$, the fourth moment is given by $\langle v^4 \rangle = \frac{3N}{N+2}$. The excess kurtosis follows as $\kappa_{\text{ex}} = \langle v^4 \rangle/\langle v^2 \rangle^2 - 3$.}
\begin{equation}
    \kappa_{\text{ex}} = \frac{3N}{N+2} - 3 = -\frac{6}{N+2}.
    \label{eq:exact_kurtosis}
\end{equation}
For sufficiently large $N$, this deviation scales asymptotically as $\kappa_{\text{ex}} \sim -6/N$. However, in the regime of our study ($5 \le N \le 20$), the exact form in Eq.~\eqref{eq:exact_kurtosis} highlights a significant deviation from the Gaussian value of $0$, providing a strong signal for tests sensitive to moment deviations.

\paragraph{Limitations of ECDF-based tests.}
Statistics such as the KS or AD tests quantify the
\emph{maximum} vertical separation between the empirical and target
cumulative distribution functions (CDFs).  
Because the finite-$N$ CDF converges uniformly to the Gaussian as
$N\to\infty$, this maximal discrepancy arises in the bulk rather than
in the tails and decays rapidly with~$N$.

\paragraph{Strength of higher-order diagnostics.}
The JB statistic combines the sample skewness and kurtosis and thus
responds directly to the excess kurtosis derived in
Eq.~\eqref{eq:exact_kurtosis}.  
The SW test, although formulated in terms of order statistics, is
likewise sensitive to tail departures and therefore retains power
longer than ECDF‐based tests.

\paragraph{A simple scaling heuristic.}
Detection is feasible when the absolute excess kurtosis
$\lvert\kappa_{\mathrm{ex}}\rvert$ exceeds the sampling noise of the
sample kurtosis.  Because the standard error of the sample kurtosis is
$\sqrt{24/n}$, the criterion reads
\[
   \lvert\kappa_{\mathrm{ex}}\rvert
   \;\gtrsim\;
   \delta\,\sqrt{\frac{24}{n}},
\]
where the detection threshold $\delta$ is defined below.
Substituting $\kappa_{\mathrm{ex}}$ from
Eq.~\eqref{eq:exact_kurtosis} gives the practical rule
\begin{equation}
    n \;\gtrsim\; \frac{2}{3}\,\delta^{2}\,(N+2)^{2},
    \label{eq:scaling_law}
\end{equation}
with
$\delta = z_{1-\alpha/2}+z_{1-\beta}$,
so that the desired two-sided significance level $\alpha$ and power
$1-\beta$ appear through the standard-normal quantiles $z_p$.
For the common choice $\alpha=0.05$ and target power $50\%$
($\beta=0.5$), one obtains $\delta\approx1.96$, hence
$n\approx2.6\,(N+2)^{2}$.
At \(N=5\) this predicts \(n\approx127\), consistent with the
\(\sim44\%\) power observed at \(n=100\).
Reaching \(90\%\) power ($\beta=0.1$, $\delta\approx3.24$) requires
about \(2.7\) times more data, in line with the simulation results.

\section{Simulation Results: Statistical Power Contour Plots}

\begin{figure}[H]
    \centering 
    
     \begin{subfigure}[b]{0.48\textwidth}
        \centering
        \includegraphics[width=\textwidth]{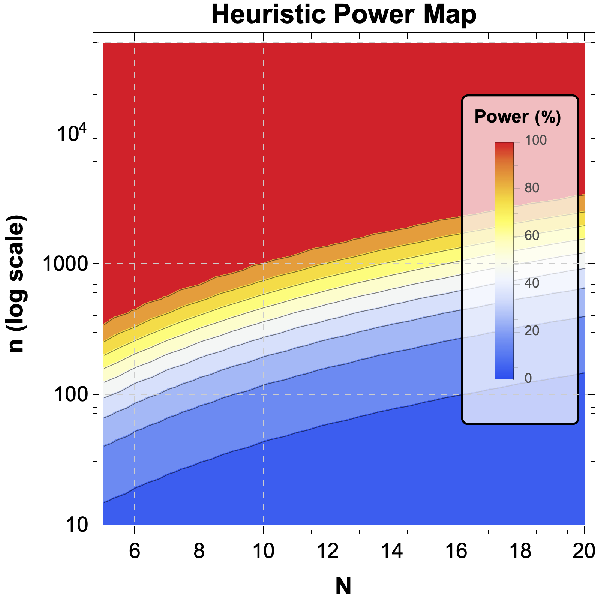}
        \caption{Heuristic Power}
        \label{fig:heuristic_power}
    \end{subfigure}
       \hfill 
    \begin{subfigure}[b]{0.48\textwidth}
        \centering
        \includegraphics[width=\textwidth]{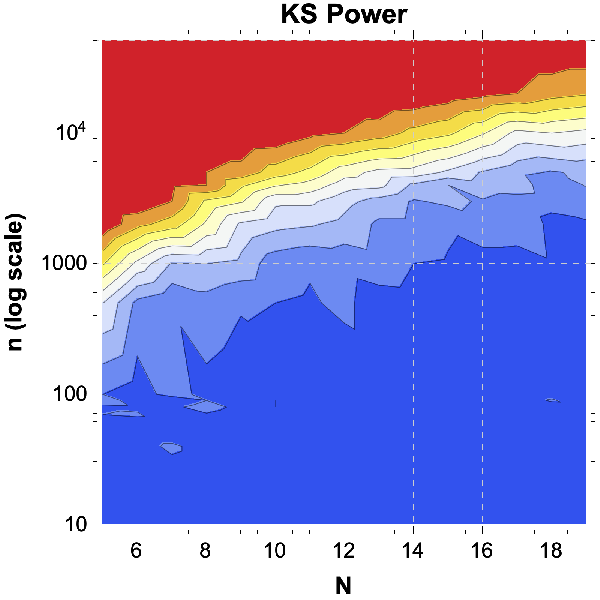}
        \caption{KS Power}
        \label{fig:ks_power}
    \end{subfigure}
    \vspace{0.5cm}     

    \begin{subfigure}[b]{0.48\textwidth}
        \centering
        \includegraphics[width=\textwidth]{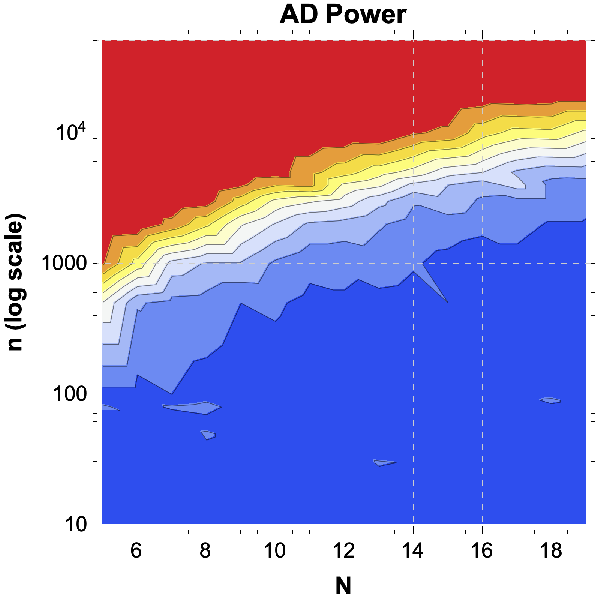}
        \caption{AD Power}
        \label{fig:ad_power}
    \end{subfigure}
    \hfill 
    \begin{subfigure}[b]{0.48\textwidth}
        \centering
        \includegraphics[width=\textwidth]{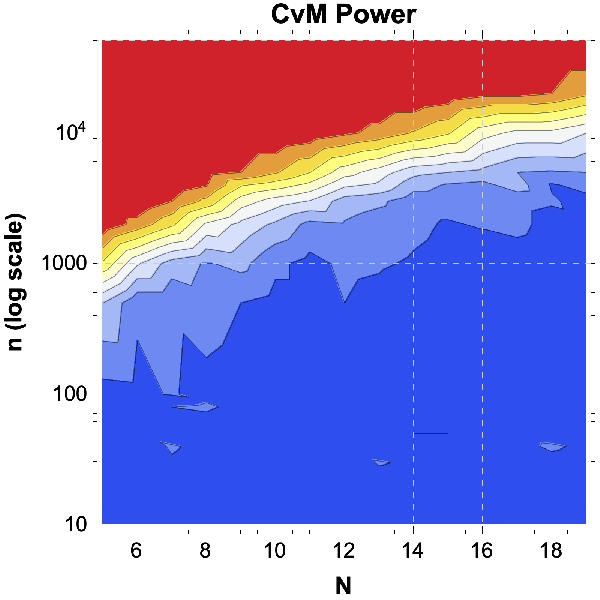}
        \caption{CvM Power}
        \label{fig:cvm_power}
    \end{subfigure}
    
    \vspace{0.5cm} 

       \begin{subfigure}[b]{0.48\textwidth}
        \centering
        \includegraphics[width=\textwidth]{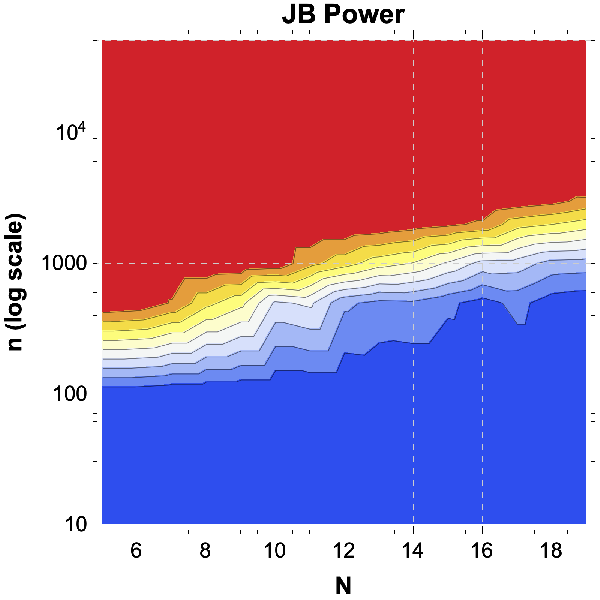}
        \caption{JB Power}
        \label{fig:jb_power}
    \end{subfigure}
    \hfill 
    \begin{subfigure}[b]{0.48\textwidth}
        \centering
        \includegraphics[width=\textwidth]{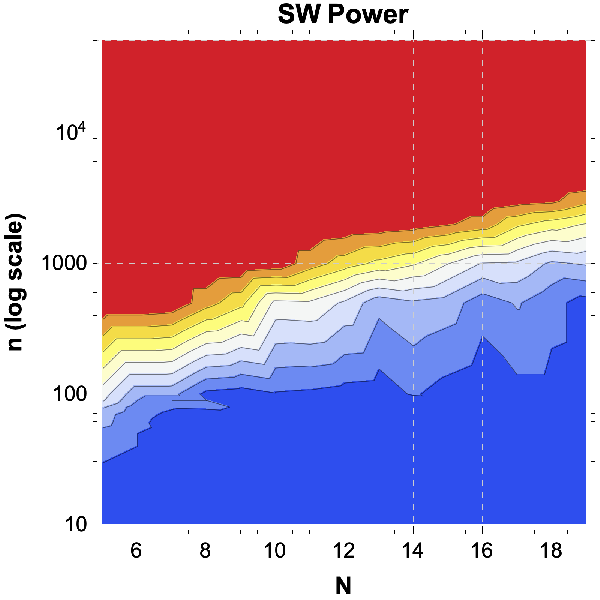}
        \caption{SW Power}
        \label{fig:sw_power}
    \end{subfigure}
    
    \caption{Statistical power contours for the heuristic approximation (a)
and five goodness-of-fit tests: Kolmogorov--Smirnov (b), Anderson--Darling (c),
Cram\'er--von Mises (d), Jarque--Bera (e), and Shapiro--Wilk (f).
The moment-sensitive tests JB and SW (e, f) retain substantially higher
power over a broad range of $(N,n)$ than the ECDF-based tests (b--d).
The sample-size axis ($n$) is plotted on a logarithmic scale.
The colour legend shown in panel (a) applies to all plots.}
    \label{fig:power_surfaces}
\end{figure}

The statistical power of five different goodness-of-fit tests was analyzed as a function of the distribution parameter $N$ and the sample size $n$. The results are visualized as contour plots in Figure~\ref{fig:power_surfaces}.

As shown in Figure~\ref{fig:power_surfaces}, the power contours for the Jarque--Bera test (Figure~\ref{fig:jb_power}) and the Shapiro--Wilk test (Figure~\ref{fig:sw_power}) are substantially higher overall than for the other tests.

\begin{figure}[h!]
    \centering     
    \includegraphics[width=0.8\textwidth]{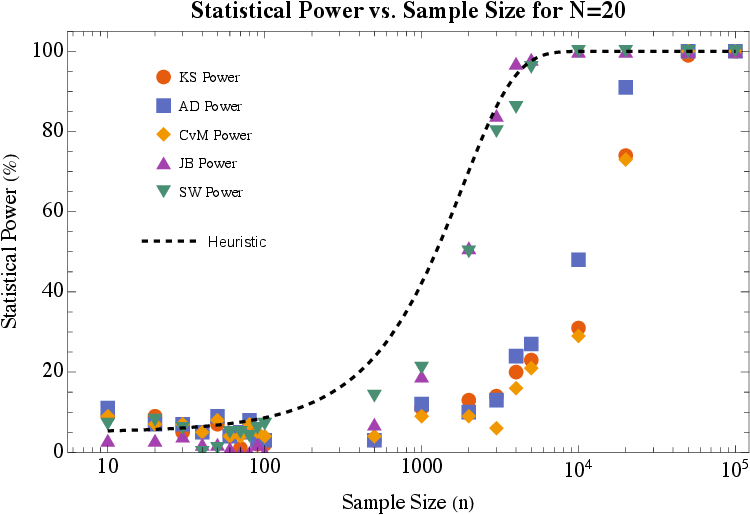}
    
\caption{Statistical power of five goodness-of-fit tests and the analytic
heuristic curve for the finite-$N$ distribution with parameter $N=20$.
The smooth ``Heuristic'' line represents the theoretical detection limit
derived from the excess-kurtosis scaling (\S\ref{sec:scaling}), against
which the simulated powers for the KS, AD, CvM, JB and SW tests are
overlaid.  ECDF-based tests (KS, AD, CvM) reach high power only for much
larger samples, whereas moment-sensitive tests (JB, SW) already achieve
$\gtrsim90\%$ power with $n\!\sim\!10^3$.  The sample-size axis ($n$) is
log-scaled.}

    \label{fig:N20_power}
\end{figure}

Figure~\ref{fig:N20_power} complements the two-parameter surfaces by
fixing \(N = 20\), where the equilibrium law is already close to Gaussian,
and sweeping the sample size \(n\) over nearly five decades.
The curves reveal two points.
First, the ECDF-based diagnostics (KS, AD, CvM) stay below
\(30\%\) power until \(n \approx \num{5000}\), confirming that shape-based
departures are hard to detect in this near-Gaussian regime.
Second, the tests sensitive to moment deviations---the Jarque--Bera
and Shapiro--Wilk tests---surpass \(95\%\) power with only a few
thousand observations, demonstrating that kurtosis provides a far
stronger signal of finite-size effects than the global CDF shape. A black dashed heuristic curve traces the theoretical detection limit.

\section{Conclusion}

We analysed how five standard goodness-of-fit tests behave on samples
drawn from the finite-\(N\) equilibrium density obtained by maximising
Havrda--Charv\'at entropy under an energy constraint
(see Eq.~\eqref{eq:finiteNpdf}).  A Monte Carlo sweep covering
\(5\!\le\!N\!\le\!19\) and \(10\!\le\!n\!\le\!5\times10^{4}\)
(plus an extended scan at \(N=20\)) yields three firm observations.

\begin{enumerate}
  \item \textbf{Type-I error is well controlled.}
        The Kolmogorov--Smirnov and Anderson--Darling tests keep their size
        close to the nominal \(5\%\) across all $(N,n)$ pairs, validating
        the simulation; the other tests deviate only mildly at the
        smallest $n$.
  \item \textbf{Kurtosis-driven tests are markedly stronger.}
        Jarque--Bera (explicitly kurtosis-based) and Shapiro--Wilk
        (order-statistic) retain high power long after KS, AD and
        Cram\'er--von Mises have lost sensitivity, confirming that finite-\(N\)
        deviations are concentrated in the excess kurtosis rather than in the
        global CDF shape.
  \item \textbf{Detectability scales as \(n\!\propto\!(N+2)^2\).}
        As \(N\) grows, all tests weaken, but at different rates:
        shape-based tests require tens of thousands of observations to
        reach 90 \% power once \(N\!\gtrsim\!15\), whereas SW and JB
        achieve the same with only a few thousand, in line with the
        heuristic scaling law of Eq.~\eqref{eq:scaling_law}.
\end{enumerate}

\paragraph{Practical takeaway.}
When non-Gaussian features are expected to be subtle, statistics that
accentuate excess kurtosis (JB) or central-shape mismatches (SW) should
be preferred; relying on KS or AD alone risks overlooking real finite-size
departures.  The heuristic contour plot in Figure~\ref{fig:heuristic_power} turns this guidance into a
ready-to-use \(n\)-versus-\(N\) benchmark.

\paragraph{Outlook.}
Future work could extend the analysis to higher spatial dimensions,
design tests tailored to compact-support alternatives, and confront the
finite-\(N\) model with empirical data---such as high-frequency financial
returns or dusty-plasma velocity distributions---to evaluate its advantage
over the Gaussian benchmark in practice.

\backmatter

\section*{Statements and Declarations}

The authors have no relevant financial or non-financial interests to disclose.\\
JWS is the sole author and was responsible for the study conception and design, analysis, drafting and revising the manuscript, and approving the final version.

\section*{Funding Declaration}

This work was partially supported by the KIST Institutional Program.

\section*{Appendix}
\begin{longtable}{rr cc ccccc}
\caption{Combined empirical power map. The table presents the validation success rate (Type-I error control) and the statistical power of five normality tests as a function of system size $N$ and sample size $n$. Data covers the full range from $n=10$ to $n=\num{50000}$.} 
\label{tab:merged_power_map} \\

\toprule
\textbf{Param.} & \textbf{Sample} & \multicolumn{2}{c}{\textbf{Validation (\%)}} & \multicolumn{5}{c}{\textbf{Statistical Power (\%)}} \\
\cmidrule(lr){3-4} \cmidrule(lr){5-9}
$\boldsymbol{N}$ & \textbf{Size} $\boldsymbol{n}$ & \textbf{KS} & \textbf{AD} & \textbf{KS} & \textbf{AD} & \textbf{CvM} & \textbf{JB} & \textbf{SW} \\
\midrule
\endfirsthead

\caption[]{Combined empirical power map (continued)} \\
\toprule
\textbf{Param.} & \textbf{Sample} & \multicolumn{2}{c}{\textbf{Validation (\%)}} & \multicolumn{5}{c}{\textbf{Statistical Power (\%)}} \\
\cmidrule(lr){3-4} \cmidrule(lr){5-9}
$\boldsymbol{N}$ & \textbf{Size} $\boldsymbol{n}$ & \textbf{KS} & \textbf{AD} & \textbf{KS} & \textbf{AD} & \textbf{CvM} & \textbf{JB} & \textbf{SW} \\
\midrule
\endhead

\midrule
\multicolumn{9}{r}{{Continued on next page}} \\
\endfoot

\bottomrule
\endlastfoot

5 & 10 & 96 & 92 & 5 & 7 & 6 & 1 & 4 \\
5 & 20 & 96 & 95 & 8 & 9 & 7 & 0 & 6 \\
5 & 30 & 98 & 97 & 4 & 4 & 4 & 0 & 10 \\
5 & 40 & 96 & 99 & 7 & 5 & 4 & 0 & 13 \\
5 & 50 & 94 & 94 & 8 & 7 & 7 & 0 & 16 \\
5 & 60 & 95 & 95 & 8 & 10 & 7 & 0 & 23 \\
5 & 70 & 95 & 93 & 10 & 9 & 9 & 0 & 24 \\
5 & 80 & 96 & 98 & 9 & 11 & 8 & 1 & 31 \\
5 & 90 & 94 & 94 & 15 & 8 & 8 & 0 & 41 \\
5 & 100 & 96 & 98 & 10 & 7 & 6 & 2 & 44 \\
5 & 500 & 94 & 98 & 40 & 49 & 30 & 100 & 100 \\
5 & 1\,000 & 91 & 96 & 71 & 91 & 68 & 100 & 100 \\
5 & 2\,000 & 96 & 95 & 98 & 100 & 98 & 100 & 100 \\
5 & 3\,000 & 97 & 98 & 100 & 100 & 100 & 100 & 100 \\
5 & 4\,000 & 94 & 95 & 100 & 100 & 100 & 100 & 100 \\
5 & 5\,000 & 98 & 97 & 100 & 100 & 100 & 100 & 100 \\
5 & 10\,000 & 98 & 95 & 100 & 100 & 100 & 100 & 100 \\
5 & 20\,000 & 96 & 97 & 100 & 100 & 100 & 100 & 100 \\
5 & 50\,000 & 96 & 93 & 100 & 100 & 100 & 100 & 100 \\
\addlinespace

6 & 10 & 98 & 98 & 4 & 3 & 3 & 2 & 1 \\
6 & 20 & 98 & 98 & 2 & 4 & 4 & 1 & 4 \\
6 & 30 & 95 & 97 & 4 & 3 & 3 & 0 & 7 \\
6 & 40 & 96 & 96 & 6 & 5 & 5 & 0 & 10 \\
6 & 50 & 96 & 94 & 4 & 5 & 8 & 0 & 10 \\
6 & 60 & 98 & 99 & 7 & 4 & 4 & 0 & 12 \\
6 & 70 & 92 & 90 & 11 & 9 & 10 & 0 & 13 \\
6 & 80 & 98 & 95 & 8 & 7 & 7 & 0 & 15 \\
6 & 90 & 98 & 97 & 5 & 7 & 6 & 0 & 20 \\
6 & 100 & 94 & 92 & 5 & 7 & 7 & 2 & 23 \\
6 & 500 & 95 & 96 & 17 & 21 & 12 & 98 & 100 \\
6 & 1\,000 & 98 & 95 & 47 & 66 & 41 & 100 & 100 \\
6 & 2\,000 & 95 & 94 & 83 & 98 & 86 & 100 & 100 \\
6 & 3\,000 & 96 & 96 & 98 & 100 & 99 & 100 & 100 \\
6 & 4\,000 & 97 & 95 & 100 & 100 & 100 & 100 & 100 \\
6 & 5\,000 & 98 & 98 & 100 & 100 & 100 & 100 & 100 \\
6 & 10\,000 & 95 & 95 & 100 & 100 & 100 & 100 & 100 \\
6 & 20\,000 & 94 & 96 & 100 & 100 & 100 & 100 & 100 \\
6 & 50\,000 & 94 & 96 & 100 & 100 & 100 & 100 & 100 \\
\addlinespace

7 & 10 & 94 & 93 & 6 & 8 & 9 & 3 & 4 \\
7 & 20 & 97 & 95 & 4 & 6 & 7 & 2 & 6 \\
7 & 30 & 93 & 92 & 8 & 9 & 9 & 0 & 3 \\
7 & 40 & 92 & 94 & 12 & 9 & 11 & 0 & 2 \\
7 & 50 & 96 & 97 & 4 & 3 & 4 & 0 & 7 \\
7 & 60 & 95 & 97 & 6 & 5 & 5 & 0 & 6 \\
7 & 70 & 98 & 95 & 3 & 8 & 4 & 0 & 8 \\
7 & 80 & 94 & 90 & 8 & 11 & 10 & 0 & 11 \\
7 & 90 & 98 & 96 & 3 & 5 & 4 & 0 & 10 \\
7 & 100 & 89 & 90 & 14 & 10 & 11 & 0 & 23 \\
7 & 500 & 93 & 93 & 11 & 17 & 12 & 89 & 98 \\
7 & 1\,000 & 97 & 96 & 29 & 36 & 25 & 100 & 100 \\
7 & 2\,000 & 94 & 92 & 75 & 87 & 72 & 100 & 100 \\
7 & 3\,000 & 94 & 95 & 90 & 98 & 90 & 100 & 100 \\
7 & 4\,000 & 97 & 96 & 92 & 100 & 97 & 100 & 100 \\
7 & 5\,000 & 98 & 97 & 98 & 100 & 99 & 100 & 100 \\
7 & 10\,000 & 93 & 93 & 100 & 100 & 100 & 100 & 100 \\
7 & 20\,000 & 95 & 95 & 100 & 100 & 100 & 100 & 100 \\
7 & 50\,000 & 97 & 95 & 100 & 100 & 100 & 100 & 100 \\
\addlinespace

8 & 10 & 92 & 91 & 8 & 9 & 8 & 6 & 5 \\
8 & 20 & 94 & 96 & 9 & 4 & 5 & 1 & 2 \\
8 & 30 & 95 & 94 & 5 & 6 & 6 & 0 & 1 \\
8 & 40 & 95 & 95 & 5 & 9 & 7 & 0 & 1 \\
8 & 50 & 95 & 91 & 8 & 11 & 9 & 0 & 1 \\
8 & 60 & 96 & 94 & 8 & 7 & 7 & 0 & 8 \\
8 & 70 & 93 & 93 & 9 & 10 & 9 & 0 & 5 \\
8 & 80 & 91 & 92 & 14 & 13 & 12 & 0 & 12 \\
8 & 90 & 96 & 94 & 10 & 9 & 9 & 0 & 10 \\
8 & 100 & 93 & 94 & 7 & 6 & 6 & 0 & 7 \\
8 & 500 & 91 & 93 & 16 & 16 & 16 & 73 & 84 \\
8 & 1\,000 & 99 & 98 & 30 & 29 & 19 & 100 & 100 \\
8 & 2\,000 & 97 & 98 & 58 & 72 & 48 & 100 & 100 \\
8 & 3\,000 & 93 & 96 & 68 & 95 & 77 & 100 & 100 \\
8 & 4\,000 & 94 & 94 & 90 & 99 & 91 & 100 & 100 \\
8 & 5\,000 & 97 & 97 & 90 & 100 & 94 & 100 & 100 \\
8 & 10\,000 & 97 & 96 & 100 & 100 & 100 & 100 & 100 \\
8 & 20\,000 & 94 & 95 & 100 & 100 & 100 & 100 & 100 \\
8 & 50\,000 & 95 & 95 & 100 & 100 & 100 & 100 & 100 \\
\addlinespace

9 & 10 & 97 & 96 & 4 & 3 & 3 & 2 & 3 \\
9 & 20 & 98 & 95 & 4 & 6 & 6 & 0 & 2 \\
9 & 30 & 92 & 94 & 9 & 7 & 8 & 1 & 3 \\
9 & 40 & 95 & 95 & 9 & 7 & 6 & 0 & 6 \\
9 & 50 & 92 & 93 & 8 & 7 & 8 & 1 & 7 \\
9 & 60 & 93 & 92 & 8 & 7 & 8 & 1 & 7 \\
9 & 70 & 97 & 97 & 4 & 3 & 3 & 0 & 7 \\
9 & 80 & 93 & 95 & 7 & 6 & 6 & 0 & 9 \\
9 & 90 & 95 & 96 & 8 & 7 & 7 & 0 & 9 \\
9 & 100 & 96 & 98 & 4 & 3 & 3 & 0 & 5 \\
9 & 500 & 95 & 96 & 11 & 10 & 10 & 63 & 73 \\
9 & 1\,000 & 97 & 94 & 27 & 29 & 23 & 99 & 99 \\
9 & 2\,000 & 97 & 96 & 38 & 51 & 37 & 100 & 100 \\
9 & 3\,000 & 97 & 97 & 61 & 75 & 55 & 100 & 100 \\
9 & 4\,000 & 93 & 95 & 72 & 92 & 71 & 100 & 100 \\
9 & 5\,000 & 96 & 96 & 86 & 97 & 90 & 100 & 100 \\
9 & 10\,000 & 92 & 93 & 100 & 100 & 100 & 100 & 100 \\
9 & 20\,000 & 93 & 93 & 100 & 100 & 100 & 100 & 100 \\
9 & 50\,000 & 93 & 95 & 100 & 100 & 100 & 100 & 100 \\
\addlinespace

10 & 10 & 95 & 95 & 5 & 5 & 5 & 4 & 6 \\
10 & 20 & 95 & 97 & 4 & 5 & 4 & 0 & 2 \\
10 & 30 & 95 & 98 & 8 & 3 & 5 & 0 & 3 \\
10 & 40 & 94 & 96 & 8 & 8 & 8 & 0 & 5 \\
10 & 50 & 91 & 95 & 10 & 6 & 7 & 1 & 3 \\
10 & 60 & 96 & 97 & 5 & 4 & 4 & 0 & 7 \\
10 & 70 & 96 & 100 & 3 & 0 & 1 & 0 & 2 \\
10 & 80 & 94 & 95 & 10 & 7 & 7 & 0 & 6 \\
10 & 90 & 91 & 92 & 10 & 10 & 8 & 0 & 6 \\
10 & 100 & 96 & 95 & 6 & 6 & 4 & 0 & 9 \\
10 & 500 & 93 & 95 & 10 & 11 & 7 & 38 & 47 \\
10 & 1\,000 & 99 & 95 & 14 & 18 & 12 & 97 & 97 \\
10 & 2\,000 & 96 & 94 & 33 & 39 & 28 & 100 & 100 \\
10 & 3\,000 & 97 & 95 & 42 & 57 & 45 & 100 & 100 \\
10 & 4\,000 & 95 & 95 & 61 & 81 & 65 & 100 & 100 \\
10 & 5\,000 & 94 & 93 & 75 & 95 & 81 & 100 & 100 \\
10 & 10\,000 & 95 & 96 & 99 & 100 & 100 & 100 & 100 \\
10 & 20\,000 & 96 & 95 & 100 & 100 & 100 & 100 & 100 \\
10 & 50\,000 & 94 & 97 & 100 & 100 & 100 & 100 & 100 \\
\addlinespace

11 & 10 & 95 & 98 & 5 & 3 & 3 & 3 & 5 \\
11 & 20 & 98 & 97 & 3 & 3 & 3 & 2 & 4 \\
11 & 30 & 96 & 95 & 4 & 5 & 4 & 1 & 5 \\
11 & 40 & 97 & 97 & 4 & 4 & 4 & 3 & 6 \\
11 & 50 & 99 & 97 & 2 & 3 & 3 & 0 & 4 \\
11 & 60 & 98 & 94 & 5 & 7 & 6 & 1 & 10 \\
11 & 70 & 95 & 94 & 4 & 6 & 6 & 0 & 1 \\
11 & 80 & 96 & 95 & 5 & 7 & 6 & 0 & 10 \\
11 & 90 & 97 & 95 & 2 & 7 & 5 & 0 & 5 \\
11 & 100 & 96 & 96 & 6 & 4 & 4 & 0 & 8 \\
11 & 500 & 96 & 94 & 9 & 8 & 7 & 42 & 48 \\
11 & 1\,000 & 94 & 97 & 11 & 12 & 7 & 83 & 85 \\
11 & 2\,000 & 94 & 96 & 31 & 28 & 17 & 100 & 100 \\
11 & 3\,000 & 95 & 95 & 42 & 51 & 39 & 100 & 100 \\
11 & 4\,000 & 94 & 95 & 56 & 83 & 57 & 100 & 100 \\
11 & 5\,000 & 96 & 96 & 67 & 83 & 70 & 100 & 100 \\
11 & 10\,000 & 94 & 96 & 93 & 100 & 97 & 100 & 100 \\
11 & 20\,000 & 93 & 94 & 100 & 100 & 100 & 100 & 100 \\
11 & 50\,000 & 97 & 94 & 100 & 100 & 100 & 100 & 100 \\
\addlinespace

12 & 10 & 99 & 98 & 1 & 2 & 1 & 5 & 6 \\
12 & 20 & 96 & 97 & 4 & 3 & 4 & 2 & 7 \\
12 & 30 & 96 & 99 & 4 & 3 & 2 & 2 & 5 \\
12 & 40 & 98 & 96 & 3 & 4 & 5 & 1 & 2 \\
12 & 50 & 99 & 97 & 1 & 5 & 4 & 0 & 4 \\
12 & 60 & 98 & 97 & 4 & 5 & 4 & 0 & 1 \\
12 & 70 & 96 & 93 & 6 & 7 & 7 & 0 & 0 \\
12 & 80 & 98 & 97 & 2 & 3 & 3 & 0 & 6 \\
12 & 90 & 95 & 94 & 5 & 6 & 7 & 0 & 4 \\
12 & 100 & 97 & 96 & 3 & 5 & 3 & 0 & 6 \\
12 & 500 & 92 & 94 & 12 & 9 & 10 & 22 & 38 \\
12 & 1\,000 & 92 & 93 & 13 & 12 & 12 & 74 & 75 \\
12 & 2\,000 & 91 & 95 & 27 & 26 & 19 & 100 & 98 \\
12 & 3\,000 & 93 & 93 & 33 & 45 & 25 & 100 & 100 \\
12 & 4\,000 & 97 & 94 & 45 & 58 & 42 & 100 & 100 \\
12 & 5\,000 & 94 & 95 & 54 & 74 & 54 & 100 & 100 \\
12 & 10\,000 & 95 & 96 & 90 & 99 & 92 & 100 & 100 \\
12 & 20\,000 & 97 & 98 & 99 & 100 & 100 & 100 & 100 \\
12 & 50\,000 & 93 & 94 & 100 & 100 & 100 & 100 & 100 \\
\addlinespace

13 & 10 & 97 & 97 & 3 & 3 & 3 & 5 & 5 \\
13 & 20 & 95 & 95 & 5 & 5 & 4 & 3 & 3 \\
13 & 30 & 92 & 91 & 8 & 11 & 11 & 1 & 4 \\
13 & 40 & 98 & 98 & 2 & 2 & 2 & 0 & 3 \\
13 & 50 & 96 & 95 & 4 & 5 & 5 & 2 & 4 \\
13 & 60 & 96 & 97 & 4 & 3 & 3 & 3 & 7 \\
13 & 70 & 95 & 96 & 8 & 6 & 6 & 0 & 2 \\
13 & 80 & 96 & 95 & 7 & 5 & 4 & 0 & 1 \\
13 & 90 & 92 & 93 & 9 & 8 & 9 & 0 & 5 \\
13 & 100 & 96 & 97 & 5 & 4 & 5 & 1 & 5 \\
13 & 500 & 96 & 94 & 5 & 7 & 7 & 17 & 23 \\
13 & 1\,000 & 96 & 95 & 16 & 15 & 11 & 64 & 67 \\
13 & 2\,000 & 98 & 95 & 18 & 21 & 15 & 98 & 97 \\
13 & 3\,000 & 94 & 93 & 26 & 33 & 22 & 100 & 100 \\
13 & 4\,000 & 96 & 94 & 43 & 52 & 38 & 100 & 100 \\
13 & 5\,000 & 97 & 94 & 47 & 64 & 45 & 100 & 100 \\
13 & 10\,000 & 93 & 92 & 85 & 100 & 87 & 100 & 100 \\
13 & 20\,000 & 93 & 93 & 100 & 100 & 100 & 100 & 100 \\
13 & 50\,000 & 92 & 98 & 100 & 100 & 100 & 100 & 100 \\
\addlinespace

14 & 10 & 91 & 92 & 9 & 10 & 10 & 4 & 2 \\
14 & 20 & 98 & 98 & 2 & 2 & 2 & 2 & 2 \\
14 & 30 & 95 & 92 & 6 & 9 & 8 & 0 & 1 \\
14 & 40 & 95 & 97 & 6 & 3 & 3 & 2 & 3 \\
14 & 50 & 93 & 92 & 8 & 8 & 10 & 0 & 4 \\
14 & 60 & 97 & 95 & 5 & 5 & 4 & 0 & 2 \\
14 & 70 & 95 & 97 & 4 & 3 & 3 & 0 & 6 \\
14 & 80 & 100 & 98 & 1 & 2 & 0 & 1 & 3 \\
14 & 90 & 97 & 94 & 6 & 8 & 7 & 0 & 5 \\
14 & 100 & 92 & 90 & 9 & 10 & 10 & 0 & 11 \\
14 & 500 & 96 & 96 & 6 & 6 & 4 & 18 & 28 \\
14 & 1\,000 & 95 & 95 & 10 & 11 & 9 & 59 & 61 \\
14 & 2\,000 & 95 & 95 & 18 & 16 & 12 & 95 & 95 \\
14 & 3\,000 & 98 & 97 & 19 & 21 & 14 & 98 & 97 \\
14 & 4\,000 & 98 & 97 & 28 & 37 & 24 & 100 & 100 \\
14 & 5\,000 & 97 & 98 & 46 & 56 & 33 & 100 & 100 \\
14 & 10\,000 & 95 & 96 & 77 & 91 & 79 & 100 & 100 \\
14 & 20\,000 & 91 & 92 & 99 & 100 & 100 & 100 & 100 \\
14 & 50\,000 & 91 & 93 & 100 & 100 & 100 & 100 & 100 \\
\addlinespace

15 & 10 & 95 & 93 & 6 & 7 & 7 & 3 & 1 \\
15 & 20 & 96 & 93 & 5 & 10 & 8 & 4 & 7 \\
15 & 30 & 99 & 99 & 1 & 2 & 1 & 5 & 2 \\
15 & 40 & 94 & 93 & 6 & 8 & 7 & 1 & 8 \\
15 & 50 & 92 & 93 & 10 & 8 & 10 & 1 & 3 \\
15 & 60 & 94 & 93 & 8 & 7 & 7 & 0 & 5 \\
15 & 70 & 94 & 94 & 5 & 7 & 7 & 0 & 4 \\
15 & 80 & 92 & 92 & 10 & 9 & 8 & 1 & 6 \\
15 & 90 & 94 & 94 & 6 & 7 & 6 & 1 & 7 \\
15 & 100 & 97 & 99 & 3 & 2 & 2 & 0 & 6 \\
15 & 500 & 95 & 93 & 9 & 10 & 10 & 12 & 25 \\
15 & 1\,000 & 97 & 96 & 9 & 7 & 7 & 48 & 50 \\
15 & 2\,000 & 96 & 96 & 12 & 15 & 9 & 92 & 91 \\
15 & 3\,000 & 96 & 95 & 22 & 27 & 14 & 100 & 98 \\
15 & 4\,000 & 95 & 94 & 20 & 30 & 19 & 100 & 100 \\
15 & 5\,000 & 94 & 95 & 38 & 44 & 30 & 100 & 100 \\
15 & 10\,000 & 100 & 97 & 72 & 88 & 71 & 100 & 100 \\
15 & 20\,000 & 94 & 95 & 96 & 100 & 97 & 100 & 100 \\
15 & 50\,000 & 93 & 93 & 100 & 100 & 100 & 100 & 100 \\
\addlinespace

16 & 10 & 96 & 97 & 4 & 4 & 4 & 2 & 4 \\
16 & 20 & 95 & 96 & 6 & 7 & 4 & 0 & 0 \\
16 & 30 & 99 & 97 & 1 & 3 & 3 & 2 & 2 \\
16 & 40 & 93 & 95 & 7 & 7 & 8 & 2 & 1 \\
16 & 50 & 95 & 96 & 4 & 4 & 4 & 1 & 2 \\
16 & 60 & 96 & 93 & 5 & 7 & 5 & 0 & 5 \\
16 & 70 & 94 & 96 & 4 & 6 & 4 & 0 & 4 \\
16 & 80 & 92 & 93 & 7 & 7 & 7 & 0 & 4 \\
16 & 90 & 99 & 97 & 1 & 2 & 2 & 1 & 2 \\
16 & 100 & 94 & 96 & 6 & 4 & 5 & 1 & 3 \\
16 & 500 & 94 & 94 & 5 & 8 & 6 & 6 & 14 \\
16 & 1\,000 & 97 & 97 & 5 & 3 & 3 & 36 & 39 \\
16 & 2\,000 & 94 & 96 & 17 & 13 & 11 & 88 & 86 \\
16 & 3\,000 & 97 & 97 & 19 & 18 & 11 & 100 & 98 \\
16 & 4\,000 & 95 & 94 & 26 & 25 & 17 & 100 & 100 \\
16 & 5\,000 & 94 & 95 & 29 & 40 & 27 & 100 & 100 \\
16 & 10\,000 & 97 & 94 & 61 & 74 & 54 & 100 & 100 \\
16 & 20\,000 & 93 & 93 & 93 & 98 & 93 & 100 & 100 \\
16 & 50\,000 & 91 & 87 & 100 & 100 & 100 & 100 & 100 \\
\addlinespace

17 & 10 & 96 & 95 & 5 & 5 & 4 & 3 & 3 \\
17 & 20 & 95 & 95 & 7 & 5 & 5 & 2 & 6 \\
17 & 30 & 95 & 95 & 5 & 5 & 4 & 2 & 3 \\
17 & 40 & 91 & 93 & 8 & 8 & 6 & 2 & 6 \\
17 & 50 & 98 & 94 & 4 & 6 & 5 & 2 & 7 \\
17 & 60 & 93 & 95 & 5 & 5 & 4 & 1 & 4 \\
17 & 70 & 96 & 96 & 5 & 4 & 5 & 1 & 4 \\
17 & 80 & 93 & 93 & 9 & 7 & 6 & 0 & 2 \\
17 & 90 & 96 & 95 & 5 & 6 & 4 & 3 & 2 \\
17 & 100 & 97 & 97 & 3 & 4 & 4 & 0 & 7 \\
17 & 500 & 95 & 97 & 7 & 5 & 6 & 13 & 21 \\
17 & 1\,000 & 96 & 98 & 5 & 7 & 4 & 35 & 38 \\
17 & 2\,000 & 95 & 95 & 16 & 13 & 13 & 74 & 72 \\
17 & 3\,000 & 95 & 95 & 14 & 17 & 13 & 96 & 94 \\
17 & 4\,000 & 91 & 89 & 28 & 34 & 25 & 99 & 99 \\
17 & 5\,000 & 94 & 93 & 23 & 28 & 20 & 100 & 100 \\
17 & 10\,000 & 95 & 98 & 47 & 66 & 47 & 100 & 100 \\
17 & 20\,000 & 92 & 93 & 90 & 99 & 93 & 100 & 100 \\
17 & 50\,000 & 96 & 96 & 100 & 100 & 100 & 100 & 100 \\
\addlinespace

18 & 10 & 97 & 98 & 3 & 3 & 4 & 5 & 6 \\
18 & 20 & 96 & 98 & 4 & 2 & 2 & 4 & 6 \\
18 & 30 & 96 & 95 & 4 & 5 & 6 & 2 & 3 \\
18 & 40 & 90 & 89 & 10 & 10 & 12 & 2 & 5 \\
18 & 50 & 99 & 98 & 1 & 2 & 2 & 3 & 4 \\
18 & 60 & 91 & 93 & 9 & 8 & 7 & 0 & 6 \\
18 & 70 & 94 & 94 & 8 & 6 & 7 & 0 & 7 \\
18 & 80 & 94 & 92 & 6 & 8 & 7 & 1 & 3 \\
18 & 90 & 92 & 90 & 11 & 12 & 10 & 0 & 6 \\
18 & 100 & 95 & 95 & 7 & 6 & 5 & 0 & 7 \\
18 & 500 & 97 & 97 & 3 & 3 & 3 & 5 & 13 \\
18 & 1\,000 & 97 & 98 & 9 & 3 & 2 & 26 & 28 \\
18 & 2\,000 & 98 & 95 & 8 & 9 & 6 & 70 & 65 \\
18 & 3\,000 & 95 & 94 & 12 & 17 & 11 & 93 & 92 \\
18 & 4\,000 & 98 & 98 & 10 & 17 & 9 & 98 & 97 \\
18 & 5\,000 & 95 & 95 & 19 & 23 & 19 & 99 & 99 \\
18 & 10\,000 & 93 & 92 & 50 & 67 & 47 & 100 & 100 \\
18 & 20\,000 & 94 & 95 & 84 & 98 & 90 & 100 & 100 \\
18 & 50\,000 & 96 & 95 & 100 & 100 & 100 & 100 & 100 \\
\addlinespace

19 & 10 & 99 & 96 & 1 & 4 & 3 & 2 & 3 \\
19 & 20 & 96 & 93 & 4 & 7 & 6 & 3 & 1 \\
19 & 30 & 94 & 96 & 5 & 4 & 3 & 1 & 6 \\
19 & 40 & 93 & 93 & 7 & 8 & 7 & 1 & 3 \\
19 & 50 & 94 & 95 & 6 & 5 & 6 & 1 & 1 \\
19 & 60 & 96 & 97 & 4 & 3 & 4 & 1 & 2 \\
19 & 70 & 94 & 94 & 7 & 6 & 6 & 4 & 6 \\
19 & 80 & 95 & 96 & 7 & 4 & 5 & 1 & 4 \\
19 & 90 & 98 & 96 & 4 & 4 & 3 & 1 & 2 \\
19 & 100 & 94 & 93 & 8 & 8 & 6 & 2 & 6 \\
19 & 500 & 95 & 93 & 5 & 7 & 6 & 3 & 6 \\
19 & 1\,000 & 98 & 94 & 9 & 9 & 7 & 25 & 31 \\
19 & 2\,000 & 92 & 93 & 9 & 9 & 8 & 65 & 59 \\
19 & 3\,000 & 97 & 97 & 14 & 13 & 7 & 87 & 83 \\
19 & 4\,000 & 94 & 93 & 21 & 16 & 13 & 99 & 94 \\
19 & 5\,000 & 96 & 96 & 22 & 24 & 20 & 100 & 99 \\
19 & 10\,000 & 92 & 91 & 47 & 57 & 40 & 100 & 100 \\
19 & 20\,000 & 91 & 92 & 81 & 98 & 82 & 100 & 100 \\
19 & 50\,000 & 97 & 94 & 100 & 100 & 100 & 100 & 100 \\
\end{longtable}


\bibliography{references}

@article{shim2021minimal,
  title={Minimal number of discrete velocities for a flow description and internal structural evolution of a shock wave},
  author={Shim, Jae Wan},
  journal={International Journal of Non-Linear Mechanics},
  volume={128},
  pages={103633},
  year={2021},
  publisher={Elsevier}
}

@article{shim2020entropy,
  title={Entropy formula of {N}-body system},
  author={Shim, Jae Wan},
  journal={Scientific Reports},
  volume={10},
  number={1},
  pages={14029},
  year={2020},
  publisher={Nature Publishing Group UK London}
}

@article{d1973tests,
  title={Tests for departure from normality. {E}mpirical results for the distributions of $b_2$ and $\sqrt{b_1}$},
  author={D'Agostino, RALPH and Pearson, Egon S},
  journal={Biometrika},
  volume={60},
  number={3},
  pages={613--622},
  year={1973},
  publisher={Oxford University Press}
}

@article{Havrda1967,
  author  = {Havrda, J. and Charvát, F.},
  title   = {Quantification Method of Classification Processes: Concept of Structural $a$-Entropy},
  journal = {Kybernetika},
  year    = {1967},
  volume  = {3},
  pages   = {30--35}
}

@article{Tsallis1988,
  author  = {Tsallis, C.},
  title   = {Possible Generalization of {Boltzmann--Gibbs} Statistics},
  journal = {Journal of Statistical Physics},
  year    = {1988},
  volume  = {52},
  pages   = {479--487}
}

@book{Tsallis2009,
  author    = {Tsallis, C.},
  title     = {Introduction to Nonextensive Statistical Mechanics: Approaching a Complex World},
  publisher = {Springer},
  address   = {New York},
  year      = {2009}
}

@article{Cho2002,
  author  = {Cho, A.},
  title   = {A Fresh Take on Disorder, or Disorderly Science?},
  journal = {Science},
  year    = {2002},
  volume  = {297},
  pages   = {1268--1269}
}

@article{Plastino1993,
  author  = {Plastino, A. R. and Plastino, A.},
  title   = {Stellar Polytropes and {T}sallis' Entropy},
  journal = {Physics Letters A},
  year    = {1993},
  volume  = {174},
  pages   = {384--386}
}

@article{Plastino1994,
  author  = {Plastino, A. R. and Plastino, A.},
  title   = {From {G}ibbs Microcanonical Ensemble to {T}sallis Generalized Canonical Distribution},
  journal = {Physics Letters A},
  year    = {1994},
  volume  = {193},
  pages   = {140--143}
}

@article{TsallisStariolo1996,
  author  = {Tsallis, C. and Stariolo, D. A.},
  title   = {Generalized Simulated Annealing},
  journal = {Physica A},
  year    = {1996},
  volume  = {233},
  pages   = {395--406}
}

@article{TsallisBemskiMendes1999,
  author  = {Tsallis, C. and Bemski, G. and Mendes, R. S.},
  title   = {Is Re‐association in Folded Proteins a Case of Nonextensivity?},
  journal = {Physics Letters A},
  year    = {1999},
  volume  = {257},
  pages   = {93--98}
}

@article{Beck2000,
  author  = {Beck, C.},
  title   = {Non‐Extensive Statistical Mechanics and Particle Spectra in Elementary Interactions},
  journal = {Physica A},
  year    = {2000},
  volume  = {286},
  pages   = {164--180}
}

@article{WaltonRafelski2000,
  author  = {Walton, D. B. and Rafelski, J.},
  title   = {Equilibrium Distribution of Heavy Quarks in {Fokker--Planck} Dynamics},
  journal = {Physical Review Letters},
  year    = {2000},
  volume  = {84},
  pages   = {31}
}

@article{IonIon2001,
  author  = {Ion, D. B. and Ion, M. L. D.},
  title   = {Evidences for Nonextensivity Conjugation in Hadronic Scattering Systems},
  journal = {Physics Letters B},
  year    = {2001},
  volume  = {503},
  pages   = {263--270}
}

@article{Upadhyaya2001,
  author  = {Upadhyaya, A. and Rieu, J.-P. and Glazier, J. A. and Sawada, Y.},
  title   = {Anomalous Diffusion and Non‐{G}aussian Velocity Distribution of {\it Hydra} Cells in Cellular Aggregates},
  journal = {Physica A},
  year    = {2001},
  volume  = {293},
  pages   = {549--558}
}

@article{WeinsteinLloydTsallis2002,
  author  = {Weinstein, Y. S. and Lloyd, S. and Tsallis, C.},
  title   = {Border between Regular and Chaotic Quantum Dynamics},
  journal = {Physical Review Letters},
  year    = {2002},
  volume  = {89},
  pages   = {214101}
}

@article{Borland2002,
  author  = {Borland, L.},
  title   = {Option Pricing Formulas Based on a Non‐{G}aussian Stock Price Model},
  journal = {Physical Review Letters},
  year    = {2002},
  volume  = {89},
  pages   = {098701}
}

@article{Lutz2003,
  author  = {Lutz, E.},
  title   = {Anomalous Diffusion and {T}sallis Statistics in an Optical Lattice},
  journal = {Physical Review A},
  year    = {2003},
  volume  = {67},
  pages   = {051402}
}

@article{ReynoldsVeneziani2004,
  author  = {Reynolds, A. M. and Veneziani, M.},
  title   = {Rotational Dynamics of Turbulence and {T}sallis Statistics},
  journal = {Physics Letters A},
  year    = {2004},
  volume  = {327},
  pages   = {9--14}
}

@article{Reis2006,
  author  = {Reis, M. S. and Amaral, V. S. and Sarthour, R. S. and Oliveira, I. S.},
  title   = {Experimental Determination of the Nonextensive Entropic Parameter {\it q}},
  journal = {Physical Review B},
  year    = {2006},
  volume  = {73},
  pages   = {092401}
}

@article{AusloosPetroni2007,
  author  = {Ausloos, M. and Petroni, F.},
  title   = {Tsallis Non‐Extensive Statistical Mechanics of El Niño Southern Oscillation Index},
  journal = {Physica A},
  year    = {2007},
  volume  = {373},
  pages   = {721--736}
}

@article{LiuGoree2008,
  author  = {Liu, B. and Goree, J.},
  title   = {Superdiffusion and Non‐{G}aussian Statistics in a Driven‐Dissipative 2{D} Dusty Plasma},
  journal = {Physical Review Letters},
  year    = {2008},
  volume  = {100},
  pages   = {055003}
}

@book{d1986goodness,
  title={Goodness-of-fit-techniques},
  editor={D'Agostino, Ralph B and Stephens, Michael A},
  year={1986},
  publisher={Marcel Dekker},
  address={New York}
  
}

@article{jarque1987,
  title={A test for normality of observations and regression residuals},
  author={Jarque, Carlos M and Bera, Anil K},
  journal={International statistical review/{R}evue internationale de statistique},
  pages={163--172},
  year={1987},
  publisher={JSTOR}
}

@book{lehmann2005,
  title={Testing statistical hypotheses},
  author={Lehmann, Erich Leo and Romano, Joseph P},
  year={2005},
  publisher={Springer},
  address={New York}
}

@article{razali2011,
  title   = {Power Comparisons of {Shapiro--Wilk, Kolmogorov--Smirnov, Lilliefors and Anderson--Darling} Tests},
  author  = {Razali, Nornadiah Mohd and Wah, Yap Bee},
  journal = {Journal of Statistical Modeling and Analytics},
  volume  = {2},
  number  = {1},
  pages   = {21--33},
  year    = {2011}
}

@article{shapiro1965,
  title={An analysis of variance test for normality (complete samples)},
  author={Shapiro, Samuel Sanford and Wilk, Martin B},
  journal={Biometrika},
  volume={52},
  number={3-4},
  pages={591--611},
  year={1965},
  publisher={Oxford University Press}
}

\end{document}